\documentstyle[12pt,a4]{article}

\begin{document}

\begin{center}
{\large \bf {NUCLEAR MOMENTS AND ELECTRON DENSITY FUNCTIONALS IN
ATOMS}}  \\

\vspace{0.5cm}

 R. L. PAVLOV$^{1}$, P. P. RAYCHEV$^{1}$, V.P. GARISTOV$^{1}$, \\
 M. DIMITROVA-IVANOVICH$^{2}$  and  J. MARUANI$^{3}$
\medskip

$^{1}$\thinspace{\it Institute for Nuclear Research and Nuclear 
Energy, \\
Bulgarian Academy of Sciences, 72 Tsarigradsko Chaussee, \\
1784 Sofia, Bulgaria - ropavlov@inrne.acad.bg}\\

\medskip

$^{2}$\thinspace{\it Institute of Solid State Physics, Bulgarian
Academy of Sciences,\\
72 Tsarigradsko Chaussee, 1784 Sofia, Bulgaria -
dimaria@phys.bas.bg}\\

\medskip
$^{3}$\thinspace {\it Laboratoire de Chimie Physique, CNRS and
UPMC,}\\
{\it 11 rue Pierre et Marie Curie, 75005 Paris, France}\\
{\it maruani@ccr.jussieu.fr}
\end{center}

\noindent ABSTRACT: An electron  density  functional approach for
the calculation of the nuclear multipole moments is presented.
The electronic matrix elements entering the experimentally
observed hyperfine electron-nucleus interaction constants in atoms
are expressed  in terms of the electron density functionals of
the charge or  spin distributions. In principle, the construction
of the charge or spin distribution density functionals can be
obtained by means of every relativistic or  non-relativistic
quantum-mechanical or DFT method. The electronic matrix  elements
for all the electronic operators of  the  existing  Hamiltonians
of hyperfine electron-nucleus  interactions are considered.\\


\noindent PACS number(s): 31.30.-i, 31.30.Gs, 33.25.Fs, 31.20.Sy.

\medskip

\section{INTRODUCTION}

The hyperfine structure in the spectra of the many-electron
systems is a result of the interaction of the electronic and
nuclear multipole (electric and magnetic) moments. This
interaction leads to the superfine splitting of the electronic
spectrum, the rotational spectrum of molecules and the electron
paramagnetic resonance lines as well \cite{Abragam61, Abragam83}.
The energy of the hyperfine electron-nucleus interaction is
determined by the finite nuclear size, the nuclear spin
orientation and the nuclear magnetic moments space distribution.
This energy is much smaller than the energy of the interactions
in an electronic system and must be considered separately and
independently for each fine structure component.

The transitions between neighbouring levels of the fine and
hyperfine structures can be measured by means of highly sensitive
radioscopic methods, for example, microwave absorption,
electron-spin resonance (ESR), nuclear magnetic resonance (NMR),
atomic and molecular beams.

These experimental methods \cite{Abragam61, Abragam83} can give
us information about many properties and effects of nuclei
connected with nuclear magnetism, for example: the mechanism and
time of spin relaxation, the width of the nuclear spin resonance
lines and the indirect interaction between nuclear spins. Using
the methods of NMR we can experimentally determine the magnetic
nuclear moments. There is no any experimental method for direct
measurement of the electric nuclear moments and especially the
quadrupole momentum. They can be determined with great accuracy
from the hyperfine splitting in electronic and molecular spectra
\cite{Ramsey}-\cite{Lucken}. For some heavy elements there are
some good muonic values of the quadrupole momentum.

In the analysis of the multipole electron-nuclear interactions
the Hamiltonian is represented as a sum of products of the
electronic and nuclear operators. These operators can be written
down explicitly and can be obtained by using the perturbation
approach by expressing the interaction Hamiltonian in series in
powers of the ratio  of nuclear and electronic   coordinates
\cite{Ramsey,Schw,Mah}, as well as by a direct introduction of
all the multipole interactions \cite{Landau,Sob}.

The operators of the magnetic electron-nuclear multipole
interactions can be determined also by expanding  the Hamiltonian
in series in powers of the constant of the fine structure which
leads to  some relativistic corrections in the main Hamiltonian
\cite{McWeeny65}-\cite{Maruani}.

Usually the experimental hyperfine multipole interaction
constants are represented as a product of the matrix elements of
the corresponding electronic and nuclear operators
\cite{Abragam61,Ramsey,Schw,Mah}. The diagonal matrix elements of
the  nuclear operators correspond to the multipole moments
(magnetic dipole, electric quadrupole, magnetic octopole nuclear
moments)\cite{Abragam61,Schw,Blatt}. In addition to the diagonal
electronic matrix elements, the non-diagonal matrix elements must
be considered as well, because of some important second order
effects. The non-diagonal matrix elements are also important in
the analysis of the magnetic transitions induced by the external
magnetic fields \cite{Ramsey}-\cite{Lucken}. Using the
theoretical interpretation of the multipole electric and magnetic
moments through  the experimental values of the hyperfine
interaction constants and the calculated multipole electronic
operators matrix elements, one can obtain the values of the
electric and magnetic nuclear moments. The calculation  of the
matrix elements of the multipole magnetic operators is essential
in determining the relative contribution of the magnetic
interactions and gives us the possibility  to estimate the role
of the quadrupole interaction in a hyperfine structure. Different
transitions in the resonance spectra make it possible to extract
the  quadrupole interaction constant by means of the analysis of
the  magnetic hyperfine sub-term or by means of the analysis of
the deviation from the interval  rules  produced by a magnetic
dipole electron-nuclear interaction. The correct
description of  the hyperfine  constants needs the inclusion of
the relativistic corrections \cite{Pernp01}.

In this paper we present the electronic matrix  elements for all
electronic operators and  express  them  in  terms of the density
functional of charge or  spin distributions. Thus for including
the relativistic  effects one must take the  corresponding density
functional of relativistic corrections, for instance, the
functional given in paper \cite{Pavlov1,Pavlov2}.

\medskip

\section{ELECTRON-NUCLEAR OPERATORS OF THE HYPERFINE
  INTERACTIONS}

\subsection{Multipole Interaction Operators}

The operators of electron-nuclear multipole interactions can be
written down \cite{Schw,Mah}
\begin{equation} \label{eq:2.1}
\hat{\mathbf{H}} = \sum_{k}\hat{\mathbf{H}}_{k}=
\sum_{k}\hat{\mathbf{Q}}^{k}\cdot\hat{\mathbf{F}}^{k}=
\sum_{k}\sum_{q}(-1)^{q}Q_{q}^{k}F^{k}_{-q},
\end{equation}       
where $\hat{\mathbf{H}}_{k}$ is the Hamiltonians of the electric
($k$-even) and of the magnetic ($k$-odd) interactions of the
nuclei with the electronic shell. Here $\hat{\mathbf{Q}}^{k}$ and
$\hat{\mathbf{F}}^{k}$ are irreducible tensor operators of $k$
rank with components $Q_{q}^{k}$ and $F^{k}_{q}$, $q=(-k, -k+1,
..., k-1, k)$. These operators act in the spaces of the total
nuclear moment $\mathbf{I}$ and of the total electronic moment
${\mathbf{J}}$ correspondingly.

The electronic operators $\hat{\mathbf{F}}^{k}$ in the
interaction Hamiltonian  expressed in series in powers of
 space coordinates are represented in the form
\begin{equation} \label{eq:2.2}
  \hat{\mathbf{F}}^{k}=\left\{
\begin{array}{ll}
 -\sum_{i}r_{i}^{-(k+1)}C^{k}(\theta_{i},\varphi_{i}), &
 (k - \mbox{even})   \\ [2ex]
\sum_{i}\mu_{e}\nabla_{i}
r_{i}^{-(k+1)}C^{k}(\theta_{i},\varphi_{i})({\frac{2}{k}}
{\mathbf{L}}(i)- 2{\mathbf{S}}(i)), & (k - \mbox{odd})
\end{array}\right.,
\end{equation}    
 where $\mu_{e}$ is the Bohr's magneton, ${\mathbf{L}}(i)$ and
 ${\mathbf{S}}(i)$ are
 the orbital and spin angular momenta of the electron with number
 $i$, ${\mathbf{r}}_{i} = (r_{i},\theta_{i},\varphi_{i})$ is its
 radius-vector and $C^{k}(\theta_{i},\varphi_{i})$ are the normalized
 spherical functions.

 The diagonal matrix elements
of the  $Q^{k}_{0}$  component of the nuclear operator
$\hat{\mathbf{Q}}^{k}$ in the state $I$, $M_{I}=I$, is the
$2^{k}$ total nuclear electric (even $k$) or the nuclear magnetic
(odd $k$) moments
\begin{eqnarray} \label{eq:2.3}
\langle I\,I|Q_{0}^k|I\,I\rangle &=& \frac{1}{\sqrt{2I+1}}\; C_{I
\ \;0 \ \;I}^{I \ k \ I} \;<I \Vert \hat{\mathbf{Q}}^k \Vert I>=
\nonumber  \\[2ex]
&=&\left\{
\begin{array}{ll}
M_k & (k - \mbox{odd}) \\[1ex]
Q_k & (k - \mbox{even})
\end{array}\right.,
\end{eqnarray}   
 where $C_{I \ \;0 \ \;I}^{I \ k \ I}$ are the Clebsch-Gordon 
coefficients
 and $<I \Vert \hat{\mathbf{Q}}^k \Vert I>$ is the reduced matrix
 element of the operator $\hat{\mathbf{Q}}^{k}$. The matrix
 elements with $k = 1,2,3$ correspond to the nuclear moments 
\cite{Schw,
 Blatt}
 $$
 M_{1}=\mu, \ \ \ Q_{2}= \frac{Q}{2}, \ \ \ M_{3}=-\Omega,
 $$
 where $\mu$ is the magnetic dipole, $Q$ is the electric quadrupole,
 and $\Omega$ is the magnetic octopole nuclear moments.

 Let us consider the case when the external field can be neglected. 
Then
in the first approximation of the perturbation theory the energy
of the hyperfine interactions is determined by the matrix elements
of the Hamiltonian $\hat{\mathbf{H}}$ Eq.(1) which are diagonal
with respect to $\mathbf{J}$, $\mathbf{I}$ and total moment
${\mathbf{F}=\mathbf{I+J}}$
\begin{equation} \label{eq:2.4}
E^{(1)}_{J}(F)= \;<\beta I \alpha J F
M_{F}|\hat{\mathbf{H}}|\beta I \alpha J F M_{F}>.
\end{equation}      
Here $M_{F}$ is the value of the $z$- projection of the total
atomic moment, $\alpha $ and $\beta $\ are the additional quantum
numbers of the electronic and nuclear systems correspondingly.

Using Racah formalism \cite{Racah} we can rewrite the expression
$E_{F}^{(1)}$ in the form
\begin{eqnarray} \label{eq:2.5}
E^{(1)}_{J}(F) &=& \sum_{k}(-1)^{I+J-F}W(II JJ;kF)<\alpha
J \Vert \hat{\mathbf{F}}^{k}\Vert\alpha J><\beta I\Vert
\hat{\mathbf{Q}}^{k}\Vert\beta I> \nonumber \\
&=& \sum_{k}A_{k}M(IJ;F;k),   
\end{eqnarray}
where
$$
A_{k}= \;<\alpha JJ|F_{0}^{k}|\alpha JJ><\beta
II|Q_{0}^{k}|\beta II>;      
$$
$$
M(IJ;F;k) = (-1)^{I+J-F}{\frac{1}{a_{k}}}W(IIJJ;kF),
$$      
$$
a_{k}={\frac{1}{\sqrt{(2J+1)(2I+1)}}}\;C^{J \ k \ J}_{J \ \;0 \
\;J}\;C^{I \ k \ I}_{I \ \;0 \ \;I},
$$     
and  $W(IIJJ;kF)$ are the Racah coefficients.

The first three hyperfine interaction constants \ $A_{k}$ \ are
connected with the experimental values of the dipole magnetic (A),
the quadrupole  electric (B) and the octopole  magnetic  (C)
interaction constants as follow
$$
A_{1}=IJA, \ \ \ A_{2}=1/4B, \ \ \ A_{3}=C
$$
Calculating the electronic matrix elements in Eq.(\ref{eq:2.5})
and using the  experimental values  for $E^{(1)}_{J}(F)$  and
$A_{k}$ we may obtain the nuclear moments $\mu$, $Q$  and
$\Omega$.

 The second order perturbation term
\begin{equation} \label{eq:2.6}
E^{(2)}_{J}(F)=\sum_{J'}{\frac{|<\beta I\alpha JF
M_{F}|\hat{\mathbf{H}}|\beta I \alpha J'F
M_{F}>|^{2}}{E_{J}-E_{J'}}},
\end{equation}
where $E_{J}$ is the energy of the fine structure level $J$,
becomes important in the case when the distance between hyperfine
structure levels is of the  same order as the energy intervals of the
fine structure.

In this case the constant corresponding to $E^{(2)}_{J}(F)$ in 
Eq.(\ref{eq:2.5})
must be added, as  a correction to the hyperfine interaction constant
 $A_{k}$ .

\medskip

\subsection{Breit-Pauli Electron-Nucleus Operators}

The Hamiltonian $\hat{\mathbf{H}}_{en}$ of the electron-nuclear
interactions obtained by means of  the expansion of the
electronic  system Hamiltonian in series in powers of the fine
structure constant includes three operators of  the effective
Breit-Pauli Hamiltonian \cite{McWeeny65}-\cite{Maruani}
\begin{equation} \label{eq:2.7}
{\hat{\mathbf{H}}}_{en} = \hat{\mathbf{H}}^{LI}_{en} +
\hat{\mathbf{H}}^{SI}_{dip} + \hat{\mathbf{H}}_{cont}^{SI}.
\end{equation}   

The electron-nuclear dipole-dipole interaction has the following
form
\begin{equation} \label{eq:2.8}
\hat{\mathbf{H}}^{SI}_{dip} = g_{0}\alpha g \alpha_{p}\sum_{i}
\left\{3[{\mathbf{S}} (i) \cdot
{\mathbf{r}}_{i}][{\mathbf{I}}\cdot{\mathbf{r}}_{i}]-
r^{2}_{i}{\mathbf{S}}(i) \cdot{\mathbf{I}}\right\}r^{-5}_{i}.
\end{equation}   
Here $g_{0}$ is the $g$-factor of the free electron, $\alpha =
1/2c$ is a half of the fine structure constant (in atomic units),
$g$ is the $g$-factor of the nucleus and $\alpha_{p}=1/2m_{p}c$.
The magnetic interaction between  the electron angular momentum
and the nuclear moment has the form
\begin{equation} \label{eq:2.9}
\hat{\mathbf{H}}^{LI}_{en}= 2\alpha g \alpha_{p}
\sum_{i}r^{-3}_{i}  {\mathbf{L}}(i) \cdot {\mathbf{I}}.
\end{equation}   

The operator of the electron-nuclear contact interaction is
\begin{equation} \label{eq:2.10}
\hat{\mathbf{H}}^{SI}_{cont} = \frac{8 \pi}{3} g_{0} \alpha g
\alpha_{p}\sum_{i} \delta ({\mathbf{r}}_{i}) {\mathbf{S}}(i)\cdot
{\mathbf{I}}.
\end{equation}   
It is straitforward to point out that the direct inclusion of
the hyperfine interactions of the nuclear momenta and the electronic
system leads to the same results as the Hamiltonian
${\hat{\mathbf{H}}}_{en}$ in  Eq.(\ref{eq:2.7}) \cite{Landau,Sob}.

\medskip

\section{MATRIX ELEMENTS OF THE ELECTRON OPERATORS}

 In order to make the matrix elements of the electronic operators 
suitable
 for  calculations we have to recouple them from the
 $LM_{L}SM_{S}JM_{J}$ coupling  scheme  to the $LM_{L}SM_{S}$
 coupling scheme. Since we consider the hyperfine splitting
 separately for each level of the fine structure, applying
 Racah techniques  \cite{Racah} we  express the  matrix elements
 $<\alpha LSJ|\hat{\mathbf{F}}^{k}|\alpha LSJ'>$ and
 $<\alpha LSJ|\hat{\mathbf{H}}_{en}|\alpha LSJ'>$
 directly through  the diagonal matrix elements of
 $\hat{\mathbf{F}}^{k}$ and  $\hat{\mathbf{H}}_{en}$
 in the $LM_{L}SM_{S}$ coupling scheme.
 Thus, along with the diagonal in $J$ matrix elements
 determining the  $A_{k}$ constants,we have to calculate the
 non-diagonal in $J$  matrix elements determining $A^{(2)}_{k}$
 values. It must be pointed out that the  non-diagonal  in $J$  
matrix elements
 are very important in  the  estimation of the transition 
probabilities between the hyperfine
 levels \cite{Abragam83,Kopf}.

\medskip

\subsection{Electric Multipole Operators}

 The symmetrical tensor operators $(\hat{\mathbf{F}}^{el})^{k}$
 Eq.(\ref{eq:2.2}) with the components \\
$(F^{el})_{q}^{k}=\sum_{i}-r_{i}^{-(k+1)}C_{q}^{k}(\theta_{i},
\varphi_{i})$ \ \ act only on the space variables,assuming
the coupling  scheme  corresponding to
 $\mathbf{J}= \mathbf{L}+\mathbf{S}$. Then  following the
Racah technique \cite{Racah}, and  using the diagonal matrix
elements
 $(\hat{\mathbf{F}}^{el})^{k}$ in the $LM_{L}SM_{S}$ coupling  scheme
 we may express the electronic  matrix elements of 
Eq.(\ref{eq:2.6})as
\begin{eqnarray}   \label{eq:3.11}
&& <\alpha LSJ|F^{k}_{q}|\alpha LSJ'> \; = (-1)^{S-L-J+k+1}
 {\sqrt{(2J+1)(2J'+1)}}  \nonumber \\
&& \times \ \ \ W(LLJJ';kS)  <\alpha
LS|\sum_{i}-r_{i}^{-(k+1)}C_{q}^{k}(\theta _{i},
\varphi_{i})|\alpha LS>.  
\end{eqnarray}
Taking  into account that the operators
$r_{i}^{-(k+1)}C_{q}^{k}(\theta_{i}, \varphi_{i})$ are
operators  of multiplication the diagonal matrix elements can be
expressed by means of the local density functional of charge
 distribution $\rho(KK|{\mathbf{r}})= \rho(K|{\mathbf{r}})$ in state
 $K \equiv |\alpha LM_{L}SM_{S}>$ in the form
 \begin{equation}  \label{eq:3.12}
 <\alpha LS|\sum_{i}r_{i}^{-(k+1)}C_{q}^{k}(\theta_{i}, \varphi
 _{i})|\alpha LS> \; = \int r^{-(k+1)}C_{q}^{k}(\theta, \varphi)
 \rho(K|{\mathbf{r}})d{\mathbf{r}}.
\end{equation}   

Up to now there is no evidence  of the existence of the electric
electron-nucleus interactions with $k > 2$. The quadrupole
interactions $(k=2)$ can be observed only in  the case of
degenerate nuclear  states and nuclear spin $I \geq 1$
\cite{Abragam61}. In our approach in the case of the quadrupole
interaction the electronic system may posses an arbitrary total
moment. As a matter of  fact in  the case  under consideration
 the density charge distribution
does not depend on  the electronic system degeneration.
        Calculation of the  matrix elements of  the
electronic operator $ $ of the quadrupole interaction is  equivalent 
[??]  to
the calculation  the  matrix elements of the  gradient of the
electrostatic field created  from  the electrons placed in the
centre  of the nucleus \cite{Abragam61},\cite{Ramsey}-\cite{Lucken},
\cite{Mah}.

\medskip

 \subsection{Magnetic Multipole Operators}

   For each value of $k$  the operators Eq.(\ref{eq:2.2}) of the
magnetic multipole interactions  are the sum of two operators.
The first term acts only on the space variables of the electronic
system
\begin{equation} \label{eq:3.13}
(\hat{\mathbf{F}}_{1}^{mg})^{k}=
{\frac{2\mu_{e}}{k}}\sum_{i}[\nabla_{i}(r_{i}^{-(k+1)}C^{k}
(\theta_{i}, \varphi_{i}))]\cdot{\mathbf{L}}(i).
\end{equation} 
In  terms of contravariant and covariant cyclic components this
operator may be written  as
\begin{equation} \label{eq:3.14}
(\hat{\mathbf{F}}_{1}^{mg})^{k}=
{\frac{2\mu_{e}}{k}}\sum_{i}\sum^{1}_{m=-1}[\nabla_{i}(r_{i}^{-
(k+1)}C^{k}
(\theta_{i}, \varphi_{i}))]^{m}L_{m}^{1}(i).  
\end{equation}
 The contravariant component of the vector in the middle brackets,
 corresponding to the component
 $C^{k}_{q}(\theta_{i}, \varphi_{i})$ of the normalized spherical
  function $C^{k}(\theta_{i}, \varphi_{i})$, can be written as
\begin{eqnarray}  \label{eq:3.15}
[\nabla_{i}(r_{i}^{-(k+1)}C_{q}^{k}(\theta_{i},\varphi_{i}))]^{m}
&=&\sqrt{\frac{4\pi}{2k+1}}[\nabla_{i}(r_{i}^{-(k+1)}
Y_{kq}(\theta_{i},\varphi_{i}))]^{m} \\ \nonumber &=&
\sqrt{4\pi(k+1)}[r_{i}^{-(k+2)}
{\mathbf{Y}}_{kq}^{(k+1)}(\theta_{i},\varphi_{i})]^{m} \\
\nonumber &=& \sqrt{(k+1)(2k+3)}r_{i}^{-(k+2)}C^{k+1 \ \;1 \
\;k}_{q-m \ m \ q}C^{k+1}_{q-m}(\theta_{i},\varphi_{i}),
\end{eqnarray}   
where ${\mathbf{Y}}_{kq}^{k+1}(\theta,\varphi)$ are spherical
vectors \cite{Varsh}, and $C^{k+1 \ \;1 \ \;k}_{q-m \ m \ q}$ are
Clebsch-Gordon coefficients.
     The scalar product in the right side of the expression
Eq.(\ref{eq:3.14}), corresponding to the component
$C_{q}^{k}(\theta_{i},\varphi_{i})$ takes the form
\begin{eqnarray} \label{eq:3.16}
&&
\sum_{m}[\nabla_{i}(r_{i}^{-
(k+1)}C_{q}^{k}(\theta_{i},\varphi_{i}))]^{m}
L_{m}^{1}(i) = \\ \nonumber
&&
\sqrt{(k+1)(2k+3)}r_{i}^{-(k+2)}[C^{k+1}(i) \times
{\mathbf{L}}(i)]^{k}_{q},    
\end{eqnarray}
where
$$ [C^{k+1}(i) \times {\mathbf{L}}(i)]^{k}_{q} \;=
\;\sum_{m}C^{k+1 \ \;1 \ \;k}_{q-m \ m \ q} \;
C^{k+1}_{q-m}(\theta_{i},\varphi_{i})L_{m}^{1}(i).
$$
Hence, the operator $(\hat{\mathbf{F}}^{mg}_{1})^{k}$  is
an irreducible tensor operator of $k-th$ rank. Since the
operators $(\hat{\mathbf{F}}^{mg}_{1})^{k}$
 act only on the space part of  the electronic system,
 for the matrix elements of the
 $q$  component $(F^{mg}_{1})^{k}_{q}$, we obtain
 \begin{eqnarray} \label{eq:3.17}
&& <\alpha LSJ|(F_{1}^{mq})_{q}^{k}| \alpha LSJ'> \;=
(-1)^{S-L-J+k+1}   \nonumber \\
&& \times \ \ \
{\frac{2\mu_{e}}{k}}{\sqrt{(2J+1)(2J'+1)(k+1)(2k+3)}}
W(LLJJ';kS)\sum_{m}C^{k+1 \ \;1 \ \;k}_{q-m \ m \ q}
\nonumber \\
&& \times \ \ \ <\alpha LS|\sum_{i}r_{i}^{-(k+2)}[C^{k+1}(i)\times
{\mathbf{L}}(i)]^{k}_{q}|\alpha LS>.
\end{eqnarray} 
 We  can represent  the sum  in Eq.(\ref{eq:3.17}) as
\begin{eqnarray} \label{eq:3.18}
&& \sum_{m}C^{k+1 \ \;1 \ \;k}_{q-m \ m \ q}<\alpha L
S|\sum_{i}r_{i}^{-(k+2)}C^{k+1}_{q-m}(\theta_{i}, \varphi_{i})
L^{1}_{m}(i)|\alpha LS> \nonumber \\
&& = \ \sum_{m}C^{k+1 \ \;1 \ \;k}_{q-m \ m \ q}\int
\limits_{{\mathbf{r}}'_{1}={\mathbf{r}}_{1}}r_{1}^{-(k+2)}C_{q-
m}^{k+1}
(\theta_{1},\varphi_{1})L^{1}_{m}(1)\rho(K|{\mathbf{r}}_{1};
{\mathbf{r}}'_{1})d{\mathbf{r}}_{1},
\end{eqnarray}  
 where $\rho(K|\mathbf{r};\mathbf{r}')$ is the non-local charge 
distribution
 density functional in \\
 $K \equiv|\alpha LM_{L}SM_{S}>$ state. In Eq.(18) the density 
functional
  does not depend on  the summing indices $m$ and $q$.

    The second operator of the magnetic multipole interactions acts
 on the space coordinates, as well as on the spin components of the
 electronic system
\begin{equation} \label{eq:3.19}
(\hat{\mathbf{F}}_{2}^{mg})^{k}= 2\mu_{e}
\sum_{i}\nabla_{i}[r_{i}^{-(k+1)}C^{k}
(\theta_{i},\varphi_{i})]\cdot{\mathbf{S}}(i).
\end{equation}   
Now  we  obtain
\begin{equation} \label{eq:3.20}
(\hat{\mathbf{F}}_{2}^{mg})^{k} = 2\mu_{e}
\sqrt{(k+1)(2k+3)}\sum_{i}r_{i}^{-(k+2)}[C^{k+1}(i)\times
{\mathbf{S}}^{1}(i)]^{k},
\end{equation}   

 The $q$ component of the tensor product in Eq.(\ref{eq:3.20})
 has the form
 \begin{equation}   \label{eq:3.21}
 [C^{k+1}(i)\times {\mathbf{S}}(i)]_{q}^{k}= \sum_{m}C^{k+1 \ \;1 \
 \;k}_{q-m \ m \ q}\;C^{k+1}_{q-m}
 (\theta_{i},\varphi_{i})S^{1}_{m}(i),
 \end{equation}  
 where the spin moment acts on the spin subsystem.
 Since both operators $C^{k+1}(i)$ and ${\mathbf{S}}(i)$ act on
 two different subsystems they commute , and
the matrix elements  of the components $(F_{2}^{mg})^{k}_{q}$ can
be expressed as
\begin{eqnarray} \label{eq:3.22}
&& <\alpha L S J|(F_{2}^{mg})^{k}_{q}|\alpha L S J'> \;= 2\mu_{e}
\sqrt{(k+1)(2k+1)(2k+3)(2J+1)(2J'+1)}   \nonumber
\\ [1ex]
&& \times \ \ \ \left\{
\begin{array}{ccc}
L & L & k+1 \\
S & S & 1 \\
J & J' & k
\end{array} \right\}\sum_{m}C^{k+1 \ \;1 \ \;k}_{q-m \ m \ q}
\nonumber \\ [1ex] && \times \ \ \ <\alpha L
S|\sum_{i}r_{i}^{-(k+2)}C^{k+1}_{q-m}
(\theta_{i},\varphi_{i})S^{1}_{m}(i)|\alpha L S>.
\end{eqnarray}  
Applying the  equations Eqs.(\ref{eq:5.61})-(\ref{eq:5.64}) of the
Appendix to the right hand matrix elements
 in Eq.(\ref{eq:3.22})  we  have
\begin{eqnarray} \label{eq:3.23}
&& <\alpha LM_{L}SM_{S}|\sum_{i}r_{i}^{-(k+2)}C^{k+1}_{q-m}
(\theta_{i},\varphi_{i})S^{1}_{m}(i)|\alpha LM_{L}SM'_{S}> =
\nonumber \\ &&
\int\limits_{{\mathbf{r}}'_{1}={\mathbf{r}}_{1}}r_{1}^{-(k+2)}C_{q-
m}^{k+1}
(\vartheta_{1},\varphi_{1}) q(K
K'|{\mathbf{r}}_{1};{\mathbf{r}}_{1}')_{m}^{1}d{\mathbf{r}}_{1}=
 \nonumber \\
&& <SM_{S}|S^{1}_{m}|SM'_{S}>\int r_{1}^{-(k+2)}
C_{q-m}^{k+1}(\vartheta_{1},\varphi_{1})D_{S}({\mathbf{r}}_{1})
d{\mathbf{r}}_{1},
\end{eqnarray} 
where $D_{S}({\mathbf{r}}_{1}) = q({\overline {K}} \ {\overline
{K}}|{\mathbf{r}}_{1};{\mathbf{r}}_{1})^{1}_{0}/S$ is the
normalized spin distribution density functional, $K'\equiv \mid
\alpha LM_{L}SM'_{S}>$, \ \ and \ \ ${\overline{K}} \equiv \mid
\alpha LM_{L}SS>$.

 The application of the Wigner-Eckart theorem for the matrix elements
  $<SM_{S}|S^{1}_{m}|SM'_{S}>$  yields
\begin{equation}  \label{eq:3.24}
<SM_{S}|S^{1}_{m}|SM'_{S}>=
\sqrt{\frac{3}{2}}\frac{1}{\sqrt{2S+1}}C ^{\ S\phantom{x} \,
1\phantom{x} \, S}_{M_S \, m \, M'_{S}}.
\end{equation}  
  Finally the contribution of the component
  $(F_{2}^{mg})^{k}_{q}$  of the magnetic operator
  $\hat{\mathbf{F}}_{2}^{mg}$
 for the hyperfine splitting of any spin  multiplet term
under consideration $(SM_{S})$ has the form
 \begin{eqnarray} \label{eq:3.25}
&& <\alpha L S J|(F_{2}^{mg})^{k}_{q}|\alpha L S J'> \;=
  {\sqrt{6}} \mu_{e} \frac{1}{\sqrt{2S+1}} \nonumber \\
 && \times \ \ \ \ {\sqrt{(k+1)(2k+1)(2k+3)(2J+1)(2J'+1)}} \nonumber
\\ [1ex]
&& \times \ \ \ \left\{
\begin{array}{ccc}
L & L & k+1 \\
S & S & 1 \\
J & J' & k
\end{array} \right\}\sum_{m}C^{k+1 \ \;1 \ \;k}_{q-m \ m \ q}
C ^{\ S\phantom{x} \, 1\phantom{x} \, S}_{M_{S} \, m \, M'_{S}}
\nonumber \\ [1ex] && \times \ \ \ \int r_{1}^{-(k+2)}
C_{q-m}^{k+1}(\vartheta_{1},\varphi_{1})D_{S}({\mathbf{r}}_{1})
d{\mathbf{r}}_{1}.
\end{eqnarray}  

\medskip

\subsection{Electron-Nucleus Dipole Operator}

The magnetic dipole coupling between electron and nuclear spins
Eq.(\ref{eq:2.8}) can be written down in a  form in  which the
electronic and nuclear variables are separated
\begin{equation} \label{eq:3.26}
 \hat{\mathbf{H}}^{SI}_{dip}=g_{0}\alpha g \alpha_{p} \sum_{i}
{\mathbf{S}}(i)\cdot[3{\frac{{\mathbf{r}}_{i}\bigotimes{\mathbf{r}}
_{i}}{r^{5}_{i}}}- 
{\frac{{\mathbf{n}}^{2}(i)}{r^{3}_{i}}}]\cdot{\mathbf{I}}.
\end{equation}      
Here ${\mathbf{n}}^{2}(i)$ is the unit tensor formed from the unit
vector ${\mathbf{n}}(i)={\mathbf{r}}_{i}/r_{i}$.

The sum of the one-electronic operators in Eq.(\ref{eq:3.26}) can
be written as
\begin{equation}  \label{eq:3.27}
 \hat{\mathbf{H}}^{SI}_{dip} = g_{0}\alpha
g_{}\alpha_{p}\sum_{i}r^{-3}_{
i}{\mathbf{K}}(i)\cdot{\mathbf{I}},
\end{equation}   
where
\begin{equation}  \label{eq:3.28}
{\mathbf{K}}(i)=3[{\mathbf{S}}(i)\cdot{\mathbf{n}}(i)]
{\mathbf{n}}(i)-
{\mathbf{S}}(i).   
\end{equation}
The components $K_{q}^{1}(i)$ of the axial vector
${\mathbf{K}}(i)$ can be written in the form
\begin{equation} \label{eq:3.29}
K_{q}^{1}(i)=\sum_{l}D_{ql}^{2}S^{1}_{l}(i),  
\end{equation}
where the second-rank, symmetric, traceless tensor
\begin{equation} \label{eq:3.30}
D_{ql}^{2}= 3n_{q}(i)n_{l}(i) - \delta_{ql}n_{q}(i)n_{l}(i)
\end{equation}    
is proportional to the normalized spherical function
$C^{2}(\vartheta_{i},\varphi_{i})$.

Then the components of the symmetrized vector
${\mathbf{K}}^{1}(i)$ are
\begin{equation}  \label{eq:3.31}
K^{1}_{q}(i) = C[D^{2}\otimes \hat{\mathbf{S}}^{1}]^{1}_{q} =
C\sum_{m,m'}C^{2\phantom{x} \, 1\phantom{x} 1}_{m \, m'\,
q}C^{2}_{m} (\vartheta_{i},\varphi_{i})S^{1}_{m'}(i),
\end{equation}  
where the $[D^{2}\bigotimes {\mathbf{S}}^{1}]^{1}_{q}$ are the
components of an irreducible first-rank tensor product of $D^{2}$
and ${\mathbf{S}}^{1}$. The constant $C$ can be derived from
atomic spectroscopy theory \cite{Sob} and is equal to $10^{1/2}$.

Using the  relations given above and expressing the scalar product of
${\mathbf{K}}(i)$ and ${\mathbf{I}}$ in Eq.(\ref{eq:3.27}) as a
tensor product of their symmetrized forms ${\mathbf{K}}^{1}(i)$
and ${\mathbf{I}}^{1}$, one obtains
\begin{equation}\label{eq:3.32}
 \hat{\mathbf{H}}^{SI}_{dip} = (\hat{\mathbf{F}}_{dip})^{1}\cdot
 \hat{\mathbf{Q}}^{1}=
 \sum_{q}(-1)^{q}(F_{dip})_{q}^{1} \cdot Q^{1}_{-q}.
\end{equation}
The $q$ component of the electronic operator has the form
\begin{eqnarray} \label{eq:3.33}
&&(F_{dip})_{q}^{1}= g_{0} \alpha \sum_{i}r^{-3}_{i}K^{1}_{q}(i)=
\nonumber \\
 && \sqrt{10}g_{0} \alpha
\sum_{m,m'}C^{2\phantom{x} \, 1\phantom{x} 1}_{m \, m'\, q}
\sum_{i}r^{-3}_{i}C^{2}_{m}
(\vartheta_{i,}\varphi_{i})S^{1}_{m'}(i),
\end{eqnarray}     
and the $q$ component of the nuclear operator  is
$$
 Q^{1}_{-q}= g \alpha_{p}I^{1}_{-q}.
$$
  The  $Q^{1}_{-q}$ components  can  be  obtained
by  means of the  Wigner-Eckart theorem. We have
\begin{eqnarray} \label{eq:3.34}
\langle I\,I|Q_{q}^1|I\,I\rangle &=& g \alpha_{p}
\frac{1}{\sqrt{2I+1}}\; C_{I \ \;q \ \;I}^{I \ 1 \ I}
\;<I\Vert\hat{\mathbf{Q}}^k\Vert I>
\\  \nonumber &=&
 g \alpha_{p} \; \{C_{I \ \;q \ \;I}^{I \ 1 \
I} / C_{I \ \;0 \ \;I}^{I \ k \ I}\} \; M_{1},
\end{eqnarray}   
where  $M_{1}=\mu$ is the magnetic dipole nuclear moment.

By analogy with Eq.(\ref{eq:2.5}), the  matrix  elements $<\beta I
\alpha J F M_{F}|\hat{\mathbf{H}}^{SI}_{dip}|\beta I \alpha J' F
M_{F}>$ can be written as
\begin{eqnarray} \label{eq:3.35}
 && <\beta I \alpha J F
M_{F}|\hat{\mathbf{H}}^{SI}_{dip}|\beta I \alpha J' F M_{F}>=
\nonumber \\
&& (-1)^{I+J'-F}W(II JJ';F)<\alpha
J\Vert(\hat{\mathbf{F}}_{dip})^{1}\Vert\alpha J'><\beta
I\Vert \hat{\mathbf{Q}}^{1}\Vert \beta I>.   
\end{eqnarray}

In order to calculate the  contribution of this interaction to
the  hyperfine splitting  and to the constant  of this splitting
$A_{k=1}$, we recouple the matrix  elements  $< \alpha LSJ |
(\hat{\mathbf{F}}_{dip})^{1}|\alpha LSJ'>$ from  the
$LM_{L}SM_{S}JM_{J}$ to the $LM_{L}SM_{S}$  coupling scheme.
 Since tensor $[C^{2}(\vartheta, \varphi)\otimes
{\mathbf{S}}^{1}]^{1}$  acts over
 space and spin subsystems the 9j-symbol technique  leads to
\begin{eqnarray} \label{eq:3.36}
 && <\alpha LSJ |(F_{dip})^{1}_{q}|\alpha LSJ'>
  =10 \sqrt{3}g_{0} \alpha
\sqrt{(2J+1)(2J'+1)}   \nonumber
\\ [1ex]
&& \times \ \ \ \left\{
\begin{array}{ccc}
L & L & 2 \\
S & S & 1 \\
J & J' & 1
\end{array} \right\}\sum_{m m'}C^{2 \ \;1 \ \;1}_{m \ m' \ q}
\nonumber \\ [1ex] && \times \ \ \ <\alpha L
S|\sum_{i}r_{i}^{-3}C^{2}_{m}
(\theta_{i},\varphi_{i})S^{1}_{m'}(i)|\alpha L S>.
\end{eqnarray}  

Using Eqs.(\ref{eq:5.61})-(\ref{eq:5.64}) of the Appendix we
express the matrix elements of the electronic factor in the form
\begin{eqnarray}\label{eq:3.37} \nonumber
&& \langle \alpha
LSM_{S}|\sum_{i}r_{i}^{-3}C^{2}_{m}(\vartheta{i},\varphi_{i})
S_{m'}^{1}(i)| \alpha LSM'_{S}\rangle = \\
&&
\int\limits_{{\mathbf{r}}'_{1}={\mathbf{r}}_{1}}r_{1}^{-3}C_{m}^{2}
(\vartheta_{1},\varphi_{1})
q(KK'|{\mathbf{r}}_{1};{\mathbf{r}}_{1}')_{m'}^{1}d{\mathbf{r}}_{1}=
 \nonumber \\
&&<SM_{S}|S^{1}_{m'}|SM'_{S}>\int r_{1}^{-3}
C_{m}^{2}(\vartheta_{1},\varphi_{1})D_{S}({\mathbf{r}}_{1})
d{\mathbf{r}}_{1},
\end{eqnarray} 

Finally for the matrix elements of the electronic dipole operator
$\hat{\mathbf{F}}^{1}_{dip}$ we obtain
\begin{eqnarray}  \label{eq:3.38}
 && < \alpha LSJ |(F_{dip})^{1}_{q}|\alpha LSJ'>
  =15 \sqrt{2}g_{0} \alpha \frac{1}{\sqrt{2S+1}}
\sqrt{(2J+1)(2J'+1)}   \nonumber
\\ [1ex]
&& \times \ \ \ \left\{
\begin{array}{ccc}
L & L & 2 \\
S & S & 1 \\
J & J' & 1
\end{array}
\right\}\sum_{m m'}C^{2 \ \;1 \ \;1}_{m \ m' \ q}
C ^{\ S\phantom{x} \, 1\phantom{x} \, S}_{M_{S} \, m \, M'_{S}}
\nonumber \\ [1ex] && \times \ \ \
\int\limits_{{\mathbf{r}}_{1}}r_{1}^{-3}
C_{m}^{2}(\vartheta_{1},\varphi_{1})D_{S}({\mathbf{r}}_{1})
d{\mathbf{r}}_{1}.
\end{eqnarray}  

One can easily see that the functional $<\alpha LSJ
|(F_{dip})^{1}_{q}|\alpha LSJ'>$ coincides, up to the constant,
with the functional of the electronic multipole operator \\
$<\alpha LSJ |(F_{2}^{mg})^{k}_{q}|\alpha LSJ'>$
(Eq.(\ref{eq:3.25})) for $k=1$.

\medskip

\subsection{Electron-Nucleus Orbital-Spin Operator}

The operator of the interaction  between the electronic orbital
momentum and the magnetic  moment of nucleus   Eq.(\ref{eq:2.9})
can be written down  as a product of  electronic and nuclear
operators and has the form
\begin{equation}\label{eq:3.39}
 \hat{\mathbf{H}}^{LI}_{en} = (\hat{\mathbf{F}}_{L})^{1}\cdot
 \hat{\mathbf{Q}}^{1}=
 \sum_{q}(-1)^{q}(F_{L})_{q}^{1} \cdot Q^{1}_{-q}.
\end{equation} 
The $q$ component of the electronic operator is
\begin{equation} \label{eq:3.40}
(F_{L})_{q}^{1}= 2 \alpha \sum_{i}r^{-3}_{i}L^{1}_{q}(i)
\end{equation}     
while the $q$ component of the nuclear operator  is
$$
 Q^{1}_{-q}= g \alpha_{p}I^{1}_{-q}.
$$

Using the Racah technique we obtain
\begin{eqnarray} \label{eq:3.41}
 && <\beta I \alpha J F
M_{F}|\hat{\mathbf{H}}^{LI}_{en}|\beta I \alpha J' F M_{F}>=
\nonumber \\
&& (-1)^{I+J'-F}W(II JJ';F)<\alpha
J\Vert(\hat{\mathbf{F}}_{L})^{1}\Vert\alpha J'><\beta
I\Vert \hat{\mathbf{Q}}^{1}\Vert \beta I>.   
\end{eqnarray}
The  matrix elements   $\langle I\,I|Q_{q}^1|I\,I\rangle$ of the
nuclear operator components $Q_{q}^1$  are again taken   from
Eq.(\ref{eq:3.34}).

The  matrix elements of the components of electronic operator
 $\hat{\mathbf{F}}^{1}_{L}$ in the $LM_{L}SM_{S}$ coupling scheme
 are
\begin{eqnarray} \label{eq:3.42}
&& <\alpha LSJ|(F_{L})_{q}^{1}|\alpha LSJ'> \; = 2 \alpha
(-1)^{S-L-J+k+1}
 {\sqrt{(2J+1)(2J'+1)}}  \nonumber \\
&& \times \ \ \ W(LLJJ';kS)  <\alpha
LS|\sum_{i}r^{-3}_{i}L^{1}_{q}(i)|\alpha LS>.  
\end{eqnarray}

And now we may represent the right side matrix elements in
Eq.(\ref{eq:3.42})in terms of the charge  distribution density
functional. Thus we obtain
\begin{equation}\label{eq:3.43}
\langle \alpha LS \mid \sum_{i}r^{-3}_{i}L^{1}_{q}(i)\mid \alpha
LS \rangle \, =
 \int\limits_ {{\mathbf{r}}_{1}^{\prime}={\mathbf{r}}_{1}}
r_{1}^{-3}L_{q}^{1}(1)\rho(KK \mid{\mathbf{r}}_{1};
{\mathbf{r}}_{1}^{\prime})d{\mathbf{r}}_{1}.
\end{equation}    
Here $\rho(KK|{\mathbf{r}}_{1};
{\mathbf{r}}_{1}^{\prime})=\rho(K|{\mathbf{r}}_{1};
{\mathbf{r}}_{1}^{\prime})$ is the non-local density
functional charge distribution.

\medskip

\subsection{Electron-Nucleus Contact Operator}

The operator of the electron-nucleus contact interactions
Eq.(\ref{eq:2.10}) can be also given as
\begin{equation}\label{eq:3.44}
 \hat{\mathbf{H}}^{SI}_{cont} = (\hat{\mathbf{F}}_{cont})^{1}\cdot
 \hat{\mathbf{Q}}^{1}=
 \sum_{q}(-1)^{q}(F_{cont})_{q}^{1} \cdot Q^{1}_{-q}.
\end{equation}   
Here  the $q$ component of the electronic operator is
\begin{equation} \label{eq:3.45}
(F_{cont})_{q}^{1}= {\frac{8 \pi}{3}} g_{0} \alpha \sum_{i}\delta
 ({\mathbf{r}}_{i})S^{1}_{q}(i)
\end{equation}     
and the $q$ component of the nuclear operator is
$$
 Q^{1}_{-q}= g \alpha_{p}I^{1}_{-q}.
$$

By  analogy with  Eq.(\ref{eq:3.41}) we  obtain
\begin{eqnarray}  \label{eq:3.46}
 && <\beta I \alpha J F
M_{F}|\hat{\mathbf{H}}^{SI}_{cont}|\beta I \alpha J' F M_{F}>=
\nonumber \\
&& (-1)^{I+J'-F}W(II JJ';F)<\alpha
J\Vert(\hat{\mathbf{F}}_{cont})^{1}\Vert\alpha J'><\beta
I\Vert\hat{\mathbf{Q}}^{1}\Vert\beta I>.   
\end{eqnarray}

The  matrix elements of the components of electronic operator
 $(\hat{\mathbf{F}}_{cont})^{1}$ in the $LM_{L}SM_{S}$ coupling 
scheme
 are
\begin{eqnarray} \label{eq:3.47}
&& <\alpha LSJ|(F_{cont})_{q}^{1}|\alpha LSJ'> \; = \frac{8 \pi}
{3} g_{0} \alpha (-1)^{S-L-J+k+1}
 {\sqrt{(2J+1)(2J'+1)}}  \nonumber \\
&& \times \ \ \ W(LLJJ';kS)  <\alpha
LS|\sum_{i}\delta({\mathbf{r}}_{i})S^{1}_{q}(i)|\alpha LS>.  
\end{eqnarray}

 Using the procedure  described in subsections 3.2 and 3.3 and the 
Appendix
equations, the electronic matrix elements can be represented
in terms of the spin distribution density  functional
\begin{eqnarray}\label{eq:3.48} \nonumber
&& \langle \alpha
LSM_{S}|\sum_{i}\delta({\mathbf{r}}_{i})S^{1}_{q}(i)|\alpha
LSM'_{S}\rangle=\left\{C^{\ S\phantom{x} \, 1\phantom{x} \,
S}_{M_S \, q \, M'_{S}}/
C^{S \, 1 \, S}_{S \, 0 \, S}\right\}  \\
&&\times \
\int\limits_{{\mathbf{r}}'_{1}={\mathbf{r}}_{1}}\delta({\mathbf{r}}
_{1})q(KK'|{\mathbf{r}}_{1};{\mathbf{r'}}_{1})^{1}_{0}
d{\mathbf{r}}_{1}= \langle SM_{S}|S^{1}_{q}|SM'_{S}\rangle
\nonumber
\\
&&\times \ \int\limits_{{\mathbf{r}}_{1}}\delta({\mathbf{r}}_{
1})D_{S}({\mathbf{r}}_{1})d{\mathbf{r}}_{1}=
\sqrt{\frac{3}{2}}\frac{1}{\sqrt{2S+1}}C ^{\ S\phantom{x} \,
1\phantom{x} \, S}_{M_S \, q \, M'_{S}}D_{S}
({\mathbf{r}})|_{{\mathbf{r}}=0},
\end{eqnarray}   
where $D_{S} ({\mathbf{r}})|_{{\mathbf{r}}=0}$ is the spin
distribution density functional at the location of the nucleus.

\medskip

\section{CONCLUSION}

In this  paper we suggest an approach for the calculation of the
nuclear multipole moments. In this approach the electronic matrix
elements entering the experimentally observed hyperfine
interaction constants are expressed  in terms of the electron
density functionals of the charge  or  spin distributions. The
matrix  elements are  presented  in a form , suitable for
numerical implementation.  These  matrix elements are determined
for  all the electronic operators entering  the Hamiltonians of
the  different   descriptions  of hyperfine electron-nuclear
interactions. Using the irreducible tensor-operators and  the
Racah  and Fano techniques as well as the density functional and
density matrix formalism, we can express these matrix  elements
as the products  of  $3j$, $6j$, or $9j$ factors  with the  space
part-charge or spin distribution density functional. In
principle, the charge or spin distribution density functionals
can be constructed using  every  relativistic or  non-relativistic
quantum-mechanical or DFT method.

The matrix elements  of  the $(\hat{\mathbf{F}}^{el})^{k}$ and
$(\hat{\mathbf{F}}_{1}^{mg})^{k}$ operators of  the Hamiltonians
$\hat{\mathbf{H}}_{k}$ Eq.(\ref{eq:2.1}) and the  operator
$(\hat{\mathbf{F}}_{L})^{1}$ of Hamiltonian
$\hat{\mathbf{H}}_{en}$ Eq.(\ref{eq:2.7}) are the charge
distribution density functionals while the matrix elements of
operator $(\hat{\mathbf{F}}_{2}^{mg})^{k}$ of the Hamiltonians
$\hat{\mathbf{H}}_{k}$ and two operators
$(\hat{\mathbf{F}}_{dip})^{1}$ and
$(\hat{\mathbf{F}}_{cont})^{1}$ of the Hamiltonian
$\hat{\mathbf{H}}_{en}$ are spin distribution density
functionals. It is important to point out that  the charge
distribution density functional does not depend on  the level of
the  fine structure, this dependence is governed by the $j$-
symbols. One may notice a similar behaviour in the case of spin
distribution density functional. But  in this case the normalized
functional remains the same within one given spin multiplet. And
here, too the fine structure level for which we consider its
hyperfine splitting within the spin multiplet is determined  by
$3j$, $6j$ or $9j$ factors Eq.(\ref{eq:3.25}),
Eq.(\ref{eq:3.38}), Eq.(\ref{eq:3.48}). Of course, the hyperfine
splitting energy depends  on $j$-symbols as well  as on the
corresponding density functional.

It is straitforward to point out that such a representation is
valid for the arbitrary degenerate or  non-degenerate states of the
electronic system and also for arbitrary  transitions induced by
an external magnetic field. Hence,in the  case of the free atoms
this approach may be successfully applied to all kinds
 of experiments determining the nuclear multipole moments.

 The charge and spin distribution density functional  may be
constructed by  means of  proper relativistic ab-initio results
(for instance,  the results of  paper \cite{Tokman}) or DFT
calculations,however of the   type which contains
 more precise relativistic corrections \cite{Pernp98}. One also may 
use the
density functional of relativistic corrections given in papers
\cite{Pavlov1,Pavlov2}.Of  course our approach gives  the possibility
 to make the non-relativistic calculations. The application of this
approach  in non-relativistic calculations  has an advantage because
 of the density functional methods  can  be used.
We have shown in subsections ${\mathbf{3.2}}$  and
${\mathbf{3.3}}$ that  the  the matrix elements Eq.(\ref{eq:3.25}) 
and  Eq.(\ref{eq:3.38})
 of the $\hat{\mathbf{F}}^{2}_{mg}$
  and  $\hat{\mathbf{F}}_{dip}$ operators  written in terms of the 
spin
  distribution  density functionals
  have the  same  analytical form in all cases. Similar
results hold for the operators sum $\hat{\mathbf{F}}_{dip} +
\hat{\mathbf{F}}_{L}$  and the operators sum
$\hat{\mathbf{F}}_{mg}=\hat{\mathbf{F}}^{1}_{mg} +
\hat{\mathbf{F}}^{2}_{mg}$ whose matrix elements, however are 
calculated only  numerically, or
 appear in  the  special case of one  electron above
 the closed shell \cite{Mah}.

\medskip

 \section{APPENDIX \\
 DENSITY FUNCTIONALS OF CHARGE AND SPIN DISTRIBUTIONS}
\medskip

\subsection{Charge Distribution Density Functionals}

The  first order density matrix of a $N$-electronic system in
state $K$ described by a wave function
$\Psi_{K}(\tau_{1},...,\tau_{N})$, eigenfunction of the operators
${\mathbf{S}}^{2}$ and $S_{z}$, has the form \cite{McWeeny65,Mest}
\begin{equation}\label{eq:5.49}
\rho(KK|\tau_{1};\tau'_{1})= N
\int\Psi_{K}(\tau_{1},\tau_{2},...,\tau_{N})
\Psi_{K}^{\ast}(\tau'_{1},\tau_{2},...,\tau_{N})\,
d\tau_{2}...d\tau_{N},
\end{equation} 
where $\tau_{i}=({\mathbf{r}}_{i},\sigma_{i})$, ${\bf r}_{i}$, \ \
$(1 \leq i \leq N)$ being the position vector and $\sigma_{i}$ the
spin variable of the $i$-th electron, and
$d\tau_{i}=d{\mathbf{r}}_{i}d\sigma_{i}$.

The corresponding first order density function  is defined by the
expression
\begin{equation}   \label{eq:5.50}
\rho(KK|\tau_{1}) = \rho(KK|\tau_{1};\tau'_{1})
|_{\tau'_{1}=\tau_{i}} = \rho(KK|\tau_{1};\tau_{1}).
\end{equation}     

The first order transition density matrix between states $K$ and
$K'$ has the form
\begin{equation} \label{eq:5.51}
\rho(KK'|\tau_{1};\tau'_{1})= N \int
\Psi_{K}(\tau_{1},\tau_{2},...,\tau_{N})
\Psi_{K'}^{\ast}(\tau'_{1},\tau_{2},...,\tau_{N}), d\tau_{2}...,
 d\tau_{N}.
\end{equation}       

For the one-density transition  function  we have a similar
expression, which follows by analogy with Eq.(\ref{eq:5.50}).

After separation of the space and spin variables, the one-density
matrix takes the form
\begin{equation} \label{eq:5.52}
\rho(\tau_{1};\tau_{1}') = \sum_{\gamma,\gamma'=\alpha,\beta}\rho
^{\gamma,\gamma'}({\mathbf{r}}_{1};{\mathbf{r}}'_{1})
\gamma(\sigma_{1})\gamma'^{\ast}(\sigma_{1}'),
\end{equation}     
where the
$\rho^{\gamma,\gamma'}({\mathbf{r}}_{1};{\mathbf{r}}'_{1})$ are
the space components of one-density matrix and the $\gamma(\sigma)
\ (\gamma = \alpha,\beta)$ are the electron spin wave-functions.

 The space density matrix of charge distribution has the form
\begin{equation}\label{eq:5.53}
\rho({\mathbf{r}}_{1};{\mathbf{r}}'_{1}) =
\int\limits_{\tau'_{1}=\tau_{i}}\rho(\tau_{1};\tau_{1}')d\tau_{1}
= \rho^{\alpha,\alpha} ({\mathbf{r}}_{1};{\mathbf{r}}'_{1}) +
\rho^{\beta,\beta} ({\mathbf{r}}_{1};{\mathbf{r}}'_{1})
\end{equation}     
and  the space density function of  charge distribution
\begin{equation}\label{eq:5.54}
\rho({\mathbf{r}}_{1}) = \rho({\mathbf{r}}_{1};{\mathbf{r}}'_{1})
 _{{\mathbf{r}}'_{1}={\mathbf{r}}_{1}} =
 \rho^{\alpha} ({\mathbf{r}}_{1}) + \rho^{\beta} ({\mathbf{r}}_{1}).
\end{equation}     
Here $\rho({\mathbf{r}}_{1})$ is local and
 $\rho({\mathbf{r}}_{1};{\mathbf{r}}'_{1})$ non-local electron 
density
functional.

 The matrix elements of a sum of identical one-particle
operators, can  be  written down as
\begin{eqnarray} \label{eq:5.55} \nonumber
\langle K | \sum_{i}\hat{{\mathbf{F}}}(i)| K' \rangle & \equiv &
\langle \Psi_{K'}(\tau_{1},...,\tau_{N})|
\sum_{i}\hat{{\mathbf{F}}}(i)|\Psi_{K} (\tau_{1},...,\tau_{N})
\rangle \\
& = &  \int\limits_{\tau'_{1} =
\tau_{i}}\hat{{\mathbf{F}}}(1)\rho(KK'| \tau_{1};\tau'_{1})d
\tau_{1}.
\end{eqnarray}     
The one particle operators $\hat{{\mathbf{F}}}(i)$  do not depend
of spin variable we have
\begin{equation} \label{eq:5.56} \nonumber
\langle K | \sum_{i}\hat{{\mathbf{F}}}(i)| K' \rangle \equiv
\int\limits_{{\mathbf{r}}'_{1}={\mathbf{r}}_{1}}\hat{{\mathbf{F}}}(1)
\rho(KK'| {\mathbf{r}}_{1};{\mathbf{r}}'_{1})d{\mathbf{r}}_{1},
\end{equation}
 and if in addition the $\hat{{\mathbf{F}}}(i)$ is the
operator of multiplication then
\begin{equation} \label{eq:5.57} \nonumber
\langle K | \sum_{i}\hat{{\mathbf{F}}}(i)| K' \rangle \equiv \int
\hat{{\mathbf{F}}}(1) \rho(KK'| {\mathbf{r}}_{1})
d{\mathbf{r}}_{1}.
\end{equation}

\medskip

\subsection{Spin Distribution Density Functionals}

In a spin eigenstate (with eigenvalues $S(S + 1)$ and $M$ ) the
spin distribution non-local density functional,  can be written
in the following form: \cite{McWeeny89,Maruani,Mest}
\begin{equation} \label{eq:5.58}
q({\mathbf{r}}_{1};{\mathbf{r}}'_{1})=1/2[\rho^{\alpha,\alpha}
({\mathbf{r}}_{1};{\mathbf{r}}'_{1})-\rho^{\beta,\beta}
({\mathbf{r}}_{1};{\mathbf{r}}'_{1})].
\end{equation}     

The spin distribution matrices
$q^{(M)}({\mathbf{r}}_{1};{\mathbf{r}}'_{1})$ for different spin
eigenstates (with $M = S, S - 1, ... , -S)$ can be expressed in
terms of the normalized spin distribution matrices
$D_{S}({\mathbf{r}}_{1};{\mathbf{r}}_{1}')$, which are
independent of $M$
\begin{equation}\label{eq:5.59}
q^{(M)}({\mathbf{r}}_{1};{\mathbf{r}}'_{1})\equiv
q(KK|{\mathbf{r}}_{1};{\mathbf{r}}'_{1})={\frac{M}{S}}q
({\overline{K}} \
{\overline{K}}|{\mathbf{r}}_{1};{\mathbf{r}}'_{1})\equiv
MD_{S}({\mathbf{r}}_{1};{\mathbf{r}}'_{1}) ,
\end{equation}    
with:
\begin{equation}\label{eq:5.60}
D_{S}({\mathbf{r}}_{1};{\mathbf{r}}'_{1})={\frac{1}{S}}q({\overline{K}
}
 \  {\overline{K}}|{\mathbf{r}}_{1};{\mathbf{r}}'_{1}).       
\end{equation}
Here $K$ is the index of the spin state corresponding to $\langle
S_{z}\rangle=M$, $\overline{K}$ is the  index of  the  state
 corresponding to the
maximal value of $M=S$.

Using the general definition for the transition spin
distribution  density functional we can  write
\cite{McWeeny65}-\cite{Maruani}
\begin{equation} \label{eq:5.61}
q(KK'|{\mathbf{r}}_{1};{\mathbf{r}}'_{1})^{1}_{m} = \int\limits_
{\sigma_{1}'=\sigma_{1}}S_{m}^{1}(1)\rho(KK'|\tau_{1};\tau_{1}')d
\sigma_{1}.    
\end{equation}
Here the $S_{m}^{1}(i)$ ($m=0,\pm 1$) are symmetrized components
of the $i$-electron spin operator ${\mathbf{S}}(i)$. Similar
formulae can be written for the transition local spin-distribution 
density
functional setting ${\mathbf{r}}_{1} = {\mathbf{r}}'_{1}$.

The spin-distribution functionals for transitions between states
$K (SM_S)$ and $K^\prime (S^\prime M_{S^\prime})$ satisfy the
relations
\begin{equation} \label{eq:5.62}
q(K K^\prime \vert \, {\bf r}_1 ; {\bf r}_1^\prime)^1_m = \left\{C
^{\ S\phantom{x} \, 1\phantom{x} \, S^\prime}_{M_S \, m \,
M_{S^\prime}}\, / C ^{S\phantom{x} \, 1\phantom{x} \,
S^\prime}_{S\phantom \, \ \overline{m} \phantom{x} \,
S^\prime}\right\} \, q(\overline{K} \overline{K}^\prime \vert \,
{\bf r}_1 ; {\bf r}_1^\prime)^1_{\overline{m}},
\end{equation}
where $C ^{\ S\phantom{x} \, 1\phantom{x} \, S^\prime}_{M_S \, m
\, M_{S^\prime}}$ are Clebsch-Gordon coefficients.
 By analogy with Eq.(\ref{eq:5.61})
one obtains
\begin{equation}\label{eq:5.63}
q(\overline{K}\overline{K}'|{\mathbf{r}}_{1};{\mathbf{r}}'_{1})^{1}_{m
}
=\int\limits_{\sigma_{1}'=\sigma_{1}}S_{m}^{1}(1)\rho
(\overline{K}\overline{K}'|\tau_{1};\tau_{1}')d\sigma_{1}.    
\end{equation}
Using the Wigner-Eckart theorem we have
\begin{eqnarray} \label{eq:5.64} \nonumber
q(K K^\prime \vert \, {\bf r}_1 ; {\bf r}_1^\prime)^1_m &=&
\left\{C ^{\ S\phantom{x} \, 1\phantom{x} \, S}_{M_S \, m \,
M'_{S}}\,/ \,C ^{S\phantom{x} \, 1\phantom{x} \, S}_{S\phantom \
\, 0 \phantom{x} \, S}\right\}  \, q(\overline{K}
\overline{K}^\prime \vert \,
{\bf r}_1 ; {\bf r}_1^\prime) \\
&=& <SM_{S} \vert S^{1}_{m} \vert SM_{S}^\prime>D_{S}({\bf r}_1 ;
{\bf r}_1^\prime),
\end{eqnarray} 
where the $S^1_m$ ($m=0, \pm 1$) are symmetrized components of the
total spin $\mathbf{S}$. Equation(\ref{eq:5.64}) is a generalization 
of  the
equation(\ref{eq:5.59}) for a transitions between different spin 
states.

\end{document}